\documentclass[aps,preprint,preprintnumbers,amssymb,amsfonts,groupedaddress,superscriptaddress]{revtex4}

\usepackage{graphicx}
\usepackage{subfigure}
\usepackage{dcolumn}
\usepackage{bm}
\usepackage{amsmath}
\usepackage{color}

\usepackage{tikz}

\usepackage{pst-node,alltt}

\newcommand{\bmath}{\begin{mathletters}}
\newcommand{\emath}{\end{mathletters}}
\newcommand{\be}{\begin{eqnarray}}
\newcommand{\ee}{\end{eqnarray}}
\newcommand{\ba}{\begin{array}}
\newcommand{\ea}{\end{array}}

\newcommand{\no}{\nonumber}

\newcommand{\bt}{\beta}
\newcommand{\de}{\delta}

\newcommand{\Ga}{\Gamma}

\newcommand{\vare} {\varepsilon}

\newcommand{\mathcalL} {\mathcal L}

\newcommand{\Tr} {{\mathrm{Tr}}}
\newcommand{\Real} {{\mathrm{Re}}}
\newcommand{\Imag} {{\mathrm{Im}}}
\newcommand{\sys} {{\mathrm{sys}}}
\newcommand{\dissp} {{\mathrm{dissp}}}
\newcommand{\decay} {{\mathrm{decay}}}
\newcommand{\trap} {{\mathrm{trap}}}
\newcommand{\opt}  {{\mathrm{opt}}}

\newcommand{\h}{\hbar}
\newcommand{\pr}{\prime}

\begin{document}
\title{Efficient Energy Transfer in Light-Harvesting Systems: Quantum-Classical Comparison, Flux Network, and Robustness Analysis}
\author{Jianlan Wu}
\affiliation{Physics Department, Zhejiang University, 38 ZheDa Road, Hangzhou, Zhejiang, 310027, China}
\affiliation{Department of Chemistry, MIT, 77 Massachusetts Ave, Cambridge, MA, 02139, USA}
\author{Fan Liu}
\affiliation{Department of Chemistry, MIT, 77 Massachusetts Ave, Cambridge, MA, 02139, USA}
\author{Jian Ma}
\affiliation{Physics Department, Zhejiang University, 38 ZheDa Road, Hangzhou, Zhejiang, 310027, China}
\author{Robert J. Silbey\footnote{dedicated to the memory of Prof. Robert J. Silbey}}
\author{Jianshu Cao\footnote{E-mail: jianshu@mit.edu.}}
\affiliation{Department of Chemistry, MIT, 77 Massachusetts Ave, Cambridge, MA, 02139, USA}
\date{\today}

\begin{abstract}
Following the calculation of optimal energy transfer in thermal environment
in our first paper (Wu \textit{et al., New J. Phys.}, 2010, \textbf{12}, 105012),
full quantum dynamics and leading-order `classical' hopping kinetics are
compared in the seven-site Fenna-Matthews-Olson (FMO) protein complex.
The difference between these two dynamic descriptions is due to
higher-order quantum corrections. 
Two thermal bath models, classical white noise (the
Haken-Strobl-Reineker model) and quantum Debye model, are
considered. In the seven-site FMO model, we observe that higher-order corrections lead to
negligible changes in the trapping time or in energy transfer
efficiency around the optimal and physiological conditions
(2\% in the HSR model and $0.1\%$ in the quantum Debye
model for the initial site at BChl 1). However, using the concept
of integrated flux, we can identify significant differences
in branching probabilities of the energy transfer
network between hopping kinetics and quantum dynamics  (26\% in the
HSR model and 32\% in the quantum Debye model for the initial site
at BChl 1). This observation indicates that the quantum coherence
can significantly change the distribution of
energy transfer pathways in the flux network with the efficiency nearly the same.
The quantum-classical comparison of the average
trapping time with the removal of the bottleneck site, BChl 4,
demonstrates the robustness of the efficient energy transfer by the
mechanism of multi-site quantum coherence. To reconcile with
the latest eight-site FMO model, the quantum-classical comparison with the
flux network analysis is summarized in the appendix.
The eight-site FMO model yields similar trapping time and network structure 
as the seven-site FMO model but leads to a more disperse distribution 
of energy transfer pathways.

\end{abstract}


\maketitle

\section{Introduction}
\label{sec01}

Natural photosynthesis is of particular interest due to its essential role as the energy source for life on earth.
In the process of biological evolution over billions of years,  photosynthetic
systems have developed optimal and robust  strategies of converting solar energy to chemical energy.
In the early stage of photosynthesis, solar energy is collected by pigments and transferred through
light-harvesting protein complexes to the reaction center for the subsequent charge-separation.
The energy conversion from photons to electrons is fast, robust, and nearly perfect in efficiency, although
the overall efficiency of photosynthesis is low. Understanding the mechanism of efficient
energy transfer in natural light-harvesting systems
can help developing low-cost and highly-efficient man-made solar energy apparatus,
including photovoltaic devices and artificial photosynthesis~\cite{Blankenship2011:Science}.

For a long time, energy transfer was considered
as an incoherent process described by hopping kinetics with Forster rate constants.
The Forster rate theory has been a prevailing theoretical technique. 
In spite of the widespread success of the Forster rate approach, recent experimental advance has shown evidence of
long-lived quantum coherence in several natural light-harvesting systems, e.g., Fenna-Matthews-Olson (FMO)~\cite{Engel2007:Nature,Panitchayangkoon2010:PNAS}
and phycocyanin 645 (PC645)~\cite{Collini2010:Nature}.
A full quantum dynamic framework becomes necessary for
studying coherent energy transfer. Many theoretical techniques have been developed to
serve this purpose~\cite{Cao1997:JCP,JLWu2010:NJP,Moix2011,Shabani2011:arXiv,Mohseni2011:arXiv2,Tanimura1989:JPSJ,
Ishizaki2009:JCP1,Ishizaki2009:PNAS,Yan2004:CPL,ChenShi2011,Palmieri2009,Coker2011,Haken1972,
Haken1973:ZPhysik,Silbey1976:ARPC,Cao2009:JPCA,Chen2011:JPCB,Plenio2008:NJP,Caruso2009:JCP,
Rebentrost2009:JPCB,Rebentrost2009:NJP,Yang2010:JCP,Yang2002:CP,Ishizaki2011,Schulten2012,Reichman2012}. 
With temporal-spatial correlation for the protein environment,
the generalized Bloch-Redfield (GBR) equation~\cite{Cao1997:JCP,JLWu2010:NJP,Shabani2011:arXiv,Mohseni2011:arXiv2,Moix2011},
the hierarchy  equation~\cite{Tanimura1989:JPSJ,Ishizaki2009:JCP1,Ishizaki2009:PNAS,Yan2004:CPL,ChenShi2011},
and other methods~\cite{Coker2011}
have successfully predicted the long-lived quantum coherent phenomenon.
Alternatively, the Haken-Strobl-Reineker (HSR) model and its generalization~\cite{Haken1972,Haken1973:ZPhysik,Silbey1976:ARPC}
have attracted much attention due to its simplicity,
although the bath noise is
classical~\cite{Cao2009:JPCA,Plenio2008:NJP,Caruso2009:JCP,Rebentrost2009:NJP,Yang2010:JCP,JLWu2010:NJP,Chen2011:JPCB}.
Recently, quantum-classical mixed methods have been also applied to the dynamics of energy transfer~\cite{
Ishizaki2011,Schulten2012,Reichman2012}.
Different theoretical methods have been
tested in the simple two-site system~\cite{Ishizaki2009:JCP1} and other complex systems~\cite{Yang2002:CP},
mainly focusing on the reliability of theoretical predictions.

However, a systematic and comprehensive investigation is still
needed to distinguish hopping kinetics and full quantum dynamics,
with the goal of quantifying nontrivial quantum effects, e.g.,
long-range quantum coherence, in a complex energy transfer network.
Throughout this paper, the long-range quantum coherence is defined
in the local site basis and excludes the contribution from the
two-state quantum dynamics. Here we will propose a quantum-classical
comparison strategy, and apply it to the seven-site FMO system with
two different descriptions of baths: the classical white noise (the
HSR model) and the quantum Debye noise. In spite of their simplicity,
quantum dynamics under these two bath models can be computed exactly and
thus can be used for a meaningful quantum-classical comparison.
Although the additional eighth site in the new eight-site FMO model can modify
quantum dynamics, the seven-site model is a good example to explore interesting
and relevant quantum phenomena, and it is also consistent with our previous paper~\cite{JLWu2010:NJP}.
In Appendix~\ref{app3}, we will present a short summary on the eight-site FMO model.
Here we will use the leading-order
kinetics: a hopping network with Fermi's golden rule rate,
which is the leading-order expansion to quantum dynamics.
With a dipole-dipole interaction between two chromophores, Fermi's
golden rule rate becomes Forster rate of energy transfer. In the
standard fashion, such hopping kinetics is considered as a
`classical' description of energy transfer. In this paper, we will
explore energy transfer in FMO using exact quantum dynamic equations
and using Fermi's golden rule rate (i.e., Forster rate) to quantify
the difference between full quantum dynamics and `classical' hopping
kinetics. This difference includes nontrivial quantum effects, e.g.,
multiple-site coherence. A systematic kinetic mapping of
quantum dynamics including high-order corrections was described in a review paper
for the classical noise~\cite{Cao2009:JPCA} and will be described for the quantum
noise in a forthcoming publication~\cite{JLWu2011:NIBA}.

In the first paper of this series~\cite{JLWu2010:NJP}, we applied
the HSR model and the GBR equation approach to optimize energy
transfer with intermediate values for various bath parameters, such
as reorganization energy, bath relaxation rate, temperature, and
spatial correlation. In particular, we found an optimal temperature
for efficient energy transfer in the seven-site FMO model.
Our results have been verified
by the hierarchic equation~\cite{Kreisbeck2010:arXiv}. The
optimization behavior has been found in other conditions such as the
spatial arrangement~\cite{Mohseni2011:arXiv2}. The site energy
optimization for the new eight-site FMO model is shown in the third
paper of this series~\cite{Moix2011}. To interpret the optimization
behavior, we have proposed the concept of trapping-free subspace,
and determined the asymptotic scalings in the weak and strong
dissipation limits~\cite{JLWu2011}. Since two-site quantum coherence
is included in Fermi's golden rule rate, `classical' hopping
kinetics can predict the optimization behavior in many
light-harvesting systems, as we will show in this paper. Therefore,
the quantum-classical comparison reported in this paper is essential
for identifying the  contribution of nontrivial quantum effects to
optimal energy transfer. Specifically, we will investigate two
quantities: the trapping time and the branching probability. The
former is directly related to the energy transfer efficiency as
shown in our first paper~\cite{JLWu2010:NJP}, whereas the latter is
a new concept constructed by directional population flux for each
two-site pair in the energy transfer
network~\cite{Cao2011:JPCB,JLWu2011:ACP,VanKampen1992}. A key
advantage of natural photosynthesis compared to its artificial
counterpart is the robustness against environmental variation and
self-protection against damages.
 Here the quantum-classical comparison is combined with
the stability analysis of energy transfer to quantify the robustness
in FMO. Our study thus provides a new approach to understand the
biological role of nontrivial quantum effects excluding the two-site
coherence, different from other theoretical papers on time-dependent
behaviors of quantum coherence and
entanglement~\cite{Sarovar2010:NaturePhys}.

The paper is organized as follows: In Sec.~\ref{sec02}, we review
the quantum dynamic framework for light-harvesting energy transfer.
In Sec.~\ref{sec02a}, we use the leading order of kinetic mapping to
define `classical' hopping kinetics, and introduce the concept of
the integrated population flux and the branching probability. In
Secs.~\ref{sec03}-\ref{sec04}, we apply the HSR model and compare
the trapping times and the branching probability in the flux
networks of FMO calculated from the classical hopping kinetics  and
full quantum dynamics. The sensitivity of parametric dependence for
the trapping time is evaluated using classical hopping kinetics. The
robustness of energy transfer is explored by removing one donor site
of FMO together with a quantum-classical comparison. In
Sec.~\ref{sec05}, we apply the Debye spectral density for the
protein environment. The trapping time and the flux network are
computed quantum mechanically using
the hierarchy equation and classically using Fermi's golden rule rate. 
In Sec.~\ref{sec06}, we conclude and discuss our results
of quantum-classical comparison and robustness analysis.
The necessary mathematical formulation is provided in Appendices~\ref{app1} and~\ref{app2}.
A short summary of the eight-site FMO model is given in Appendix~\ref{app3}
with emphasis on the quantum-classical comparison and the flux network analysis.

\section{Liouville dynamics and transfer efficiency}
\label{sec02}

In this section, we review the theoretical framework of exciton dynamics, following the same notation as
introduced in a previous review article~\cite{Cao2009:JPCA} and the first paper of this three-part series~\cite{JLWu2010:NJP}.

For each local chromophore (site) of the exciton system, a two-level truncation is reliable for the lowest electronic excitation.
In consistence with low light absorption in natural light-harvesting systems,
we consider the situation of single excitation, and
then energy transfer dynamics can be studied
in the subspace of single-excitation quantum states. Thus, we introduce
a tight-binding Hamiltonian in the site ($\{|n\rangle\}$)
representation~\cite{May2004},
\be
H = \sum_{n} \vare_n |n\rangle\langle n| + \sum_{m\neq n} J_{mn} |m\rangle\langle n|,
\label{eq02}
\ee
where $\vare_m$ is the excitation energy at chromophore site $m$ and $J_{mn}$ is the electronic coupling strength
between the $m$-th and $n$-th sites.
The system investigated in this paper is the
Fenna-Matthews-Olson (FMO) protein complex with seven bacteriochlorophyll (BChl)
sites~\cite{Ermler1994,Li1997:JMB,Vulto1998:JPCB,Cho2005:JPCB,Brixner2005:Nature,Adolphs2006:BPJ,Engel2007:Nature,Panitchayangkoon2010:PNAS}.
To be consistent, we use the particular Hamiltonian model in our first paper~\cite{JLWu2010:NJP}.
The possibility of the eighth site in FMO has been addressed recently~\cite{SchmidtamBusch2011:JPCL}, and the optimization of
energy transfer regarding the new FMO model is studied in the third paper of this series~\cite{Moix2011}.

For an exciton system, the time evolution of the reduced density matrix $\rho(t)$ is governed by
the Liouville equation~\cite{Mukamel1995,May2004,Nitzan2006,Cao2009:JPCA},
\be
\dot{\rho}(t) = -\left[\mathcalL_{\sys}+\mathcalL_{\trap}+\mathcalL_{\decay}+\mathcalL_{\dissp}\right] \rho(t).
\label{eq01}
\ee
The four Liouville superoperators $\mathcalL$ on the right-hand side of the above equation
correspond to four distinct dynamic processes, which are discussed as follows.
For an isolated system, the system Liouville superoperator $\mathcalL_{\sys}$
is given by the commutator of the system Hamiltonian, $\mathcalL_{\sys}\rho=i[H, \rho]$,
and its explicit form in the Liouville space is
\be
[\mathcalL_{\sys}]_{mn, kl} = i(H_{mk}\de_{n,l}-H_{ln}\de_{m,k}).
\label{eq01a}
\ee
For conciseness, we neglect the reduced Planck constant $\h$ throughout this paper.

The irreversible population depletion of the exciton system originates from exciton decay by the electron-hole recombination
and energy trapping at the reaction center~\cite{Cao2009:JPCA,JLWu2010:NJP}.
The Liouville superoperators of these two processes are diagonal: $[\mathcalL_{\decay}]_{mn}=k_{d; mn}=(k_{d,m}+k_{d,n})/2$,
and $[\mathcalL_{\trap}]_{mn}=k_{t; mn}=(k_{t,m}+k_{t,n})/2$, where $k_{d, n}$ and $k_{t, n}$
are phenomenological decay and trapping rate constants at site $n$, respectively.
Here  $[\mathcalL]_{mn}=[\mathcalL]_{mn, mn}$ represents the diagonal element.
In practice, we often assume a homogeneous decay process with $k_{d; n}=k_d$.
In the FMO system, BChl 3 is the trap site connecting to the reaction center,
$k_{t; n} = k_t \de_{n, 3}$, and the trapping rate is set to be $k_t=1$ ps$^{-1}$. 

In addition to the above three dynamic processes, the excitation energy transfer is modulated by fluctuations
due to the interaction between the exciton system and the protein environment. On the microscopic level,
$\mathcalL_{\dissp}$ is evaluated using the explicit system-bath Hamiltonian.
Within this description, the linearly-coupled harmonic bath, $H_{SB} = \sum_n |n\rangle\langle n| B_n$,
is widely applied, with $B_n$ the linear quantum operator of bath~\cite{Mukamel1995,May2004,Nitzan2006}.
The dissipative dynamics of system is then fully determined by the bath spectral
density $J(\omega)$.
For simplicity, we ignore the spatial correlation of bath in this paper and discuss its effect in the future.
Next we can apply quantum dynamic
methods,  e.g., the Redfield equation,
the generalized Bloch-Redfield equation~\cite{Cao1997:JCP,JLWu2010:NJP,Shabani2011:arXiv,Mohseni2011:arXiv2,Moix2011},
and the Forster equation, under various approximations.
For a Gaussian bath whose time correlation function can be represented as a linear combination of exponentially decaying functions,
the hierarchy equation approach can provide a reliable prediction of quantum dynamics~\cite{Tanimura1989:JPSJ,Ishizaki2009:JCP1,Ishizaki2009:PNAS}.

Alternatively, we can view the system-bath interaction
as a time-dependent fluctuation on the system Hamiltonian~\cite{Haken1972,Haken1973:ZPhysik,Silbey1976:ARPC,Tanimura1989:JPSJ,Cao1996:},
i.e., $H(t) = H +\de H(t)$, with $\langle \de H(t)\rangle =0$.
The dissipative dynamics can  be fully determined if all the time-averaged moments of
$\de H(t)$ are resolved, which is usually an unfeasible task. In the extremely high temperature
limit, $\de H(t)$ behaves classically and the relevant second-order moment becomes real.
One example of this approximation is the HSR model where a classical white noise,
$\langle \de \vare_m(t)\de \vare_n(0)\rangle = \Ga^\ast \de_{m,n}\de(t)$, is assumed on site energies~\cite{Haken1972,Haken1973:ZPhysik}.
The dissipation Liouville superoperator becomes diagonal in the site representation,
$\left[\mathcalL_{dissp}\right]_{mn} = (1-\de_{m,n})\Ga^\ast$, where $\Ga^\ast$ is the pure dephasing rate.
Since the HSR model can be rigorously solved, it serves as the simplest model to examine our
kinetic mapping of quantum dynamics.

A key quantity of excitation energy transfer is the energy
transfer efficiency $q$, which is the ratio of energy trapping at the reaction center.
The mathematical definition of $q$ is given by
\be
q = \int_0^\infty dt~ \Tr\left\{\mathcalL_{\trap}\rho(t)\right\} = \sum_{n} k_{t;n}\tau_{n},
\label{eq05}
\ee
where $\tau_n$ is the mean residence time at site $n$,  $\tau_n = \int_0^\infty dt \rho_n (t)$.
For an arbitrary vector $X$ in Liouville space (e.g., the density matrix $\rho$),
its trace is defined as $\Tr\{X\}=\sum_n X_{nn}$. In nature, spontaneous energy decay occurs
on the time scale of nanosecond,
much slower than the picosecond energy transfer process.
The condition of $k_d \approx 1$ ns$^{-1} \ll k_t$ allows us to simplify
the transfer efficiency to~\cite{Cao2009:JPCA,JLWu2010:NJP},
\be
q \approx \frac{1}{1+k_d\langle t\rangle},
\label{eq06}
\ee
where $\langle t\rangle = \sum_n \tau_n(k_d=0)$  is the mean first passage time to the trap state in the absence
of decay (i.e., the average trapping time).
The comparison of transfer efficiencies calculated from Eqs.~(\ref{eq05}) and (\ref{eq06})
has been examined in our first paper~\cite{JLWu2010:NJP},
and their excellent agreement over a broad range of $\Ga^\ast$ proves the reliability of Eq.~(\ref{eq06}). In this paper,
we will ignore the energy decay process and focus on the average trapping time $\langle t\rangle$.
Following the formal solution of Eq.~(\ref{eq01}), the average trapping time,
\be
\langle t \rangle = \Tr\left\{\mathcalL^{-1} \rho(0)\right\}_{k_d=0},
\label{eq07}
\ee
is determined by the Liouville superoperator $\mathcalL =\mathcalL_{\sys}+\mathcalL_{\trap}+\mathcalL_{\dissp}$ and the initial
condition $\rho(0)=\rho(t=0)$. For the FMO system, BChl 1 and BChl 6 connected to the baseplate are considered
as two initial sites for energy transfer~\cite{Ishizaki2009:PNAS}.
In our calculation, we consider two initial conditions at either BChl 1 ($\rho_1(0)=1$, initial condition I)
or BChl 6 ($\rho_6(0)=1$, initial condition II).

\section{Kinetic Mapping, Flux Network, and Branching Probability}
\label{sec02a}
\subsection{Kinetic Mapping}
\label{sub-a}

In our first paper, we have demonstrated the generality of optimal
energy transfer by the competition of quantum coherence and
bath-induced relaxation~\cite{JLWu2010:NJP}. A remaining question is
to identify  contributions of nontrivial quantum effects. To do
this, we systematically map the energy transfer process to a kinetic
process. With the Markovian approximation, the quantum kinetic
equation reads~\cite{Cao2009:JPCA}
\be
\dot{P}_m=-\sum_{n\neq m}\left(k^Q_{mn}P_m-k^Q_{nm}P_n\right)-k_{t, m}P_m,
\label{eq11new}
\ee
where $P_m=\rho_{mm}$ is the population at site $m$. The
effective quantum kinetic rate $k^Q_{mn}$ can be formally derived
following the Laplace transformation, as shown in
Appendix~\ref{app1}. In the HSR model, the kinetic mapping is solved
in a recent feature article~\cite{Cao2009:JPCA}, following an
alternative stationary approximation for quantum coherence
$\rho_{mn}(t)$. In a general quantum network, the kinetic mapping in
a rigorous non-Markovian form will be left in a forthcoming
paper~\cite{JLWu2011:NIBA}. The formal derivation in Appendices~\ref{app1} and~\ref{app2} 
is sufficient for understanding quantum-classical comparison in this paper.
Equation~(\ref{eq11new}) can be
organized into a matrix form as $\dot{P} = -(K^Q+K_t)P$, where
$[P]_m = P_m$ is the population vector. The two rate matrices, $K^Q$
and $K_t$, are defined as $[K^Q]_{m, n(\neq m)} = -k^Q_{m, n}$,
$[K^Q]_{n, n} = \sum_{m(\neq n)} k^Q_{m, n}$, and $[K_t]_{m, n} =
\de_{m, n}k_{t; n}$.
The average trapping time can be alternatively defined as 
\be
\langle t\rangle = \sum_n \tau_n = \sum_{n} [(K^Q+K_t)^{-1}P(0)]_n.
\label{eq12new}
\ee
which is exactly the same as that in Eq.~(\ref{eq07}).

In this kinetic mapping, the leading order term represents the
`classical' hopping behavior in the site basis, and higher-order
corrections represent nontrivial quantum coherent effects. In
practice, the full quantum kinetic rates $K^Q$ are difficult to
evaluate exactly whereas the leading-order hopping rates $K^C$ can
be calculated using Fermi's golden rule
expression~\cite{JLWu2011:NIBA}. For the `classical' hopping kinetics,
the rate equation in Eq.~(\ref{eq11new}) and the trapping time in
Eq.~(\ref{eq12new}) remain the same after the replacement of the
classical rate matrix $K^C$. To distinguish quantities calculated by
full quantum dynamics and by classical hopping kinetics, we denote
the quantum results by $\{\tau^Q_n,~ \langle t\rangle_Q\}$ and the
classical results by $\{\tau^C_n,~ \langle t\rangle_C\}$. The
difference between the two trapping times is attributed to
higher-order quantum corrections, e.g., multi-site quantum
coherence.

\subsection{Integrated Population Flux}
\label{sub-b}

To further reveal the difference between quantum and classical
energy transfer, we construct the flux network defined by
directional population
flows~\cite{Cao2011:JPCB,JLWu2011:ACP,VanKampen1992}. For a
classical kinetic network, the integrated population flux $F^C_{mn}$
is defined by the net population flow from site $n$ to site $m$,
\be
F^C_{mn} =&& k^C_{mn}\tau^C_n-k^C_{nm}\tau^C_m = \int_0^{\infty}
k^C_{mn} P^C_n(t) -k^C_{nm} P^C_m(t) dt.
\label{apeq01}
\ee
The quantum population flux $F^Q_{mn}$ can be similarly defined by
replacing the classical residence time $\tau^C_n$ and the hopping
rate $k^C$ with the quantum residence time $\tau^Q$ and the
effective quantum rate $k^Q$ from the kinetic mapping.
Alternatively,  we will rewrite $F^Q_{mn}$ in
terms of the coherence decay time, $\tau^Q_{mn}=\int_0^\infty dt
\rho_{mn}(t)$. As derived in detail in Appendix~\ref{app2},
the quantum integrated population flux is given by
\be
F^Q_{mn} = k^Q_{mn}\tau^Q_n-k^Q_{nm}\tau^Q_m
= 2~ \Imag\left[ J_{mn}\tau^Q_{nm}\right].
\label{app_eq020}
\ee
Equations~(\ref{apeq01}) and (\ref{app_eq020}) will be used in this
paper to calculate the population fluxes in the leading-order
hopping kinetics and in full quantum dynamics, respectively, and
their difference will reveal nontrivial quantum effects that cannot
be revealed by the average trapping time $\langle t\rangle$ or
efficiency.

\subsection{Basic Properties of Flux Network}
\label{sub-c}

The concept of the integrated population flux is important in
understanding network structure and dynamics. The classical flux was
introduced in the study of enzymatic networks. Its quantum
mechanical counterpart share many of the basic properties:

\begin{itemize}
\item
Integrated flux accounts for the net population transfer in energy
transfer processes. It vanishes for an equilibrium system because of
the detailed balance condition and is an intrinsic property of
non-equilibrium steady state (NESS) systems. Light-harvesting energy
transfer is an irreversible NESS process driven by absorbed photons
to the reaction center and is therefore characterized by the
integrated flux.
\item
The flux is a conserved quantity which is normalized to unity for
every absorbed photon. For a one-dimensional chain system, the flux
is unit for every link, both in classical kinetics and in quantum
dynamics. As a result, the integrated flux is a unique quantity to
characterize the topology of an energy transfer network and to
compare the quantum and classical flows in the network.
\item
The conservation of the integrated flux applies both globally and
locally. For any state on the network, the sum of fluxes is zero,
$\sum_{m} F_{mn} =0$, so that the sum of fluxes into the state is
identical to the sum of fluxes out of the state. As a result, we can
define the branching probability,
\be
q_{mn} =
\frac{F_{mn}}{\sum_{m^\pr, F_{m^\pr n}>0}F_{m^\pr n}},
\ee
to describe the normalized probability of $n\rightarrow m$ from the
starting point $n$. The complete set of $q_{mn}$ then characterizes
the flow pattern on the network. In the rest of this paper, the
integrated population flux $F_{mn}$ will be used interchangeably
with the branching probability $q_{mn}$.
\item
In a kinetic network, the integrated population flux $F_{mn}$ and
the residence time $\tau_n$ can be simultaneously solved by imposing
the flux conservation relationship (i.e., the flux balance
approach)~\cite{Cao2011:JPCB,JLWu2011:ACP}. This flux method can
significantly reduce the computational cost for a large-scale
network. As shown in the study of FMO in this paper, we will use the
sign and magnitude of $F_{mn}$ to quantify the reaction pathways in
an irreversible network.
\end{itemize}

\section{Quantum-Classical Comparison in the Haken-Strobl-Reineker Model}
\label{sec03}

\subsection{Trapping Time}
\label{sec03a}

For the HSR model, the quantum dynamics of the seven-site FMO system
has been solved in our first paper~\cite{JLWu2010:NJP}. Following
the kinetic mapping, the 'classical' hopping rate between sites $m$
and $n$ is given by~\cite{Hoyer2010:NJP,Cao2009:JPCA}
\be
k^C_{mn} =
k^C_{nm} = \frac{2\Ga_{mn}}{\Ga^2_{mn}+\Delta^2_{mn}}|J_{mn}|^2,
\label{eq10}
\ee
where the site energy difference is
$\Delta_{mn}=\varepsilon_m-\varepsilon_n$ and the overall dephasing
rate is $\Ga_{mn}=\Ga^\ast_{mn}+k_{t;mn}$. In Appendix~\ref{app1},
we further prove that Eq.~(\ref{eq10}) can be recovered from Fermi's
golden rule rate under a classical white noise. Next we calculate
the trapping time in both full quantum dynamics and classical
hopping kinetics. In Ref.~\cite{Cao2009:JPCA}, we have proven that
$\langle t\rangle_Q$ and $\langle t\rangle_C$ are the same for the
two-site system in the HSR model. Thus, the difference between these
two trapping times arises from multi-site quantum coherence.

In Fig.~\ref{fig01}, we plot the average quantum trapping time $\langle t\rangle_Q$
and the classical counterpart $\langle t\rangle_C$ as functions of the pure dephasing rate $\Ga^\ast$
with the two initial conditions for the seven-site FMO system.
We observe that the values of $\langle t\rangle$ computed
from the two different methods are close. For example, the relative difference between
$\langle t\rangle_Q$ and $\langle t\rangle_C$ is 2\% under the optimal condition of $\Ga^\ast_\opt=175$ cm$^{-1}$ for the initial
population at BChl 1, while the difference becomes less than 1\% under $\Ga^\ast_\opt=195$ cm$^{-1}$
for the initial population at BChl 6.
Overall, the relative trapping time difference
is always less than 10\% for $\Ga^\ast\gtrapprox 30$ cm$^{-1}$.

Our result shows that full quantum dynamics and hopping kinetics lead to
similar behaviors in the trapping time and energy transfer efficiency around the optimal condition.
In the HSR model, energy transfer in the seven-site FMO is controlled by the downhill pathway
from BChl 6 to BChl 3, which does not need long-range exchange assisted by multi-site quantum coherence~\cite{Brixner2005:Nature}.
More importantly, the trapping time is the sum of residence time at all the sites so that
the cancelation from different sites can reduce the quantum effect.
Unlike the classical white noise,
a quantum colored noise at a finite temperature
complicates the quantum-classical comparison, which will be discussed in Sec.~\ref{sec05}.

\subsection{Flux Network}
\label{sec03b}

In this subsection, we present the quantum-classical comparison of the population flux for the seven-site FMO system
in the HSR model.  For the two-site system in the HSR model,
we can prove that Eqs.~(\ref{apeq01}) and (\ref{app_eq020}) lead to the same result, consistent with
the result of the trapping time.
In the over-damped limit, i.e., $\Ga^\ast\gg |J|$, quantum dynamics reduces to classical hopping kinetics,
and the two definitions become identical for multi-site networks.
Apart from this limit, we compare the results obtained with Eqs.~(\ref{apeq01}) and (\ref{app_eq020}),
and use their difference to define the
higher-order contribution, e.g., multi-site quantum coherence in the HSR model.

Figure~\ref{fig02} presents relevant population fluxes ($>0.05$) calculated by full quantum dynamics ($F^Q_{mn}$)
and hopping kinetics ($F^C_{mn}$), with two initial conditions in FMO. For each initial condition,
its respective optimal pure dephasing rate $\Ga^\ast_{\opt}$ is used.
Both quantum and classical flux networks show two identical dominating pathways:
$1\rightarrow 2 \rightarrow 3$  (path A) and $6\rightarrow (5, 7)\rightarrow 4 \rightarrow 3$ (path B).
This result is consistent with 2D electronic spectroscopy~\cite{Brixner2005:Nature}.
With initial condition II, path A
is nearly negligible whereas with initial condition I,
path B contributes significantly with $F^Q_{34}=0.40$ ($F^C_{34}=0.56$).
As shown in the next section, the two-pathway structure in FMO helps maintain
its high efficiency even if one donor is removed.

Compared to results of the average trapping time, the quantum-classical difference
of fluxes is much larger. For the branching probability from BChl 1 to BChl 3, the weak electronic coupling
$J_{13}$ leads to a small value of $F^C_{31}=0.053$ in the hopping picture with initial condition I.
On the other hand, multi-site quantum coherence allows the long-range population transfer through
interconversion between various off-diagonal elements of the density matrix~\cite{Nitzan2006,JLWu2011:NIBA}.
This quantum tunneling effect enhances the quantum branching probability more than twice, $F^Q_{31}=0.123$.
For the entire network, we introduce the weighted relative difference between classical and
quantum branching probabilities as
\be
\chi_F = \frac{2\sum_{m\neq n} |F^Q_{mn}-F^C_{mn}|}{\sum_{m\neq n} |F^Q_{mn}+F^C_{mn}|}.
\ee
The result is $\chi_F = 26\%$ for initial condition I and $\chi_F = 7\%$ for initial condition II. Both values
are significantly larger than those for the trapping time ($\le2\%$). The comparison of population fluxes
quantifies the contribution of multi-site quantum coherence in the HSR model.

The quantum-classical comparison of branching probabilities clarifies  the structure of various energy transfer pathways.
For the first path of $1\rightarrow 2 \rightarrow 3$, our Hamiltonian shows that BChl 2 is the energy barrier along the transfer pathway.
As we discussed, the quantum tunneling effect is relevant, assisting energy transfer.
For the second pathway of $6\rightarrow (5, 7)\rightarrow 4 \rightarrow 3$,
the energy downhill structure allows quick energy transfer even in
the classical hopping picture, and quantum effect is less relevant. The weighted relative flux difference $\chi_F$
is much larger in initial condition I than that in initial condition II, consistent with energy structures of these pathways.

\section{Robustness of Energy Transfer in the Haken-Strobl-Reineker Model}
\label{sec04}

\subsection{Stability Against the Variation of System/Bath Parameters}
\label{sec04a}

The leading-order kinetic network provides a simple estimate of the parametric dependence of
the average trapping time and can explain the robustness of energy transfer via the
insensitivity to changes in dephasing rate, trapping rate, and system Hamiltonian.
Within the kinetic network, estimation based on the magnitudes of the hopping rate and
trapping rate yields the average trapping time on the order of picosecond time scale,
$\langle t\rangle \sim 10$ ps.  In comparison with the average decay rate on the nanosecond time scale,  $k_d=1$ ns$^{-1}$,
the average trapping rate is much faster and therefore the efficiency is close to one,  $q\sim 1$.
The small ratio of the trapping time and decay time, $k_d\langle t\rangle\sim 0.01$,
suggests that significant drop in energy transfer efficiency will result from a change of two-order's magnitude in the trapping time.
According to Eq.~(\ref{eq10}), changes of two-order's magnitude in  $\Ga^\ast$ and $k_t$
or changes of one-order's magnitude in the $J$  and $\Delta$ disorders are needed to produce this drop,
based on simple estimation.
Although the above simple estimation can deviate from the rigorous quantum dynamics, our method reveals
that FMO can resist a large change in dephasing rate, trapping rate, and system Hamiltonian, mainly
because of the time scale separation between the decay process ($\mathcal L_{\mathrm{decay}}$) and the other
three dynamic processes ($\mathcal L_{\mathrm{sys}}$, $\mathcal L_{\mathrm{dissp}}$, and $\mathcal L_{\mathrm{trap}}$.

\begin{table}
\begin{tabular}{|c|c|c|c|c|c|c|c|}
\hline
Site Removed& 1 & 2 &  4 & 5 & 6 & 7  & complete FMO\tabularnewline \hline
$\langle t\rangle_C(\mathrm{initial~ condition~ I})$ (ps)  &    &10.57  & 15.32 & 9.60 & 10.61 &9.38 &10.30\tabularnewline  \hline
$\langle t\rangle_Q(\mathrm{initial~ condition~ I})$ (ps)  &    & 10.32 &12.41  &9.17  &9.84 &9.13 &10.10\tabularnewline  \hline
$\langle t\rangle_C(\mathrm{initial~ condition~ II})$ (ps) & 7.47  &  7.61 & 16.12 &  7.98 & &7.78 &8.72\tabularnewline  \hline
$\langle t\rangle_Q(\mathrm{initial~ condition~ II})$ (ps) & 7.51   & 7.72 & 14.31 & 7.80 & &7.63 &8.66\tabularnewline  \hline
\end{tabular}
\caption{The quantum ($\langle t\rangle_Q$) and classical ($\langle t\rangle_C$)
trapping times of FMO with the removal of one non-trapping site
under the respective optimal dephasing rate for two initial conditions in the HSR model.
As a comparison, the results for the complete seven-site model are shown in the last column.}
\end{table}

\subsection{Robustness Against the Loss of a BChl Chromophore}
\label{sec04b}

Over the course of evolution, the FMO complex has achieved robustness such that the high
energy transfer efficiency is retained even if one or two chromophores do not function properly~\cite{Kim2010}.
We investigate this feature by taking out one non-trapping site off the seven-site
FMO network and then calculating the average trapping time under the optimal condition.
As shown in Table I, removal of BChl 4 causes a noticeable increase in the average trapping
time while removal of any other site do not have major effects and some
can even enhance the transfer efficiency.
The flux distribution in Fig.~\ref{fig02}
indicates that BChl 4 is the bottleneck of the dominant pathway $6\rightarrow (5, 7)\rightarrow 4 \rightarrow 3$,
where all fluxes in this pathway converge to site 4 before arriving at the trap site.
However, even with BChl 4 removed, the increase in the average trapping time (or equivalently the decrease in efficiency) is moderate.
This phenomenon is due to the two major pathways in the energy transfer network of FMO.
Without multiple pathways in energy transfer networks, a linear-chain system exhibits
much stronger response since the single pathway would be blocked after the removal of a donor.
In addition, the trapping time decreases 12-20\% (see Table I) in full quantum dynamics than in classical hopping kinetics,
which demonstrates that the quantum coherence can help the FMO system further resist the damage on BChl 4. Next
we plot the relation of $\langle t\rangle$ - $\Ga^\ast$ in Fig.~\ref{fig01}
using both quantum dynamics and hopping kinetics for the FMO system with BChl 4 removed.
Our results show that the quantum trapping time is consistently smaller than the classical result when BChl 4 is removed,
verifying the robustness of energy transfer due to multi-site quantum coherence.

\section{Quantum-Classical Comparison in a Quantum Non-Markovian Bath}
\label{sec05}

In a general quantum network, the leading-order hopping rate is the same as Fermi's golden rule rate,
which becomes the Forster rate if the dipole-dipole interaction $J_{mn}$ is applied.
In detail, Fermi's golden rule rate is written as
\be
k^C_{m\neq n} =&& 2 |J_{mn}|^2 \Real \int_0^\infty dt e^{i \Delta_{nm} t} e^{-[g_{mn}(t)+k_{t; mn}t]},
\label{eq13a}
\ee
and
\be
g_{mn}(t) = 2(1-c_{mn}) \int_0^\infty d\omega \frac{J(\omega)}{\omega^2} \left[\coth(\bt\omega/2)(1-\cos\omega t)+i \sin\omega t\right],
\ee
where $\bt = 1/k_BT$, $J(\omega)$ is the bath spectral density, and $c_{mn}$ is the bath spatial correlation
between sites $m$ and $n$. The negative correlation $c_{mn(\neq m)}=-1$ is used in the spin-boson model,
and  the zero spatial correlation $c_{mn}=\de_{m,n}$ is considered in this paper.
Consistent with our first paper~\cite{JLWu2010:NJP}, we use the Debye spectral density,
\be
J(\omega) = \Theta(\omega) \frac{2\lambda}{\pi}\frac{\omega D}{\omega^2+D^2},
\ee
with $\Theta(\omega)$ the step function, $\lambda$ the reorganization energy, and $D$ the Debye frequency.
The real bath spectral density of the protein environment in FMO is more complicated with high-frequency
signature of local vibrational modes. The approximate Debye spectral density, however, can
predict reasonably well for the light spectra of FMO.
Besides, quantum dynamics is extremely difficult to be calculated rigorously
for a complex bath spectral density. Therefore, we will keep the simple Debye spectral density in
this paper. Without trapping, the ratio of forward and backward hopping rate constants satisfies the detailed
balance condition, $k^C_{mn}/k^C_{nm} = \exp[-\bt (\vare_n-\vare_m)]$.
The hopping rates from Eq.~(\ref{eq13a}) are
then used to compute the trapping time and population fluxes in `classical' kinetics.

Applying the Matsubara expansion~\cite{May2004},
the bath time correlation function $C(t)$ of the Debye spectral density follows
\be
C(t)=\sum_{j=0}^\infty (f^r_j+i f^i_j)e^{-\nu_j t},
\label{eq16x}
\ee
where the zeroth decay rate is $\nu_0=D$ and all the other decay rates are the Matsubara
frequencies $\nu_{j\ge 1}=2\pi j/\bt$. The coefficients $f^r_j$ and $f^i_j$ can be
determined accordingly~\cite{Ishizaki2009:PNAS}.
For the exponentially decaying bath, quantum dynamics can  be
reliably solved by the hierarchy equation approach in
principle~\cite{Tanimura1989:JPSJ,Ishizaki2009:JCP1,Ishizaki2009:PNAS,Yan2004:CPL}.
Here we use the explicit form
shown in Ref.~\cite{Ishizaki2009:PNAS}, with the trapping Liouville superoperator $\mathcal L_{\trap}$ included
for both the reduced density matrix and auxiliary fields.
To reduce the computation cost, the high-temperature approximation,
$\coth(\bt\omega/2)\approx 2/\bt\omega$, is applied so that all the Matsubara frequencies $\nu_{j\ge 1}$
are ignored in Eq.~(\ref{eq16x}).
To be consistent, the same approximation is used in computing hopping rates.
The high-temperature approximation will not cause a significant difference in our calculation at room temperature ($T=300$ K).
Our quantum computation is truncated upto the 10th hierarchic order.
Due to instability of Liouville superoperators, $\tau^Q$ is evaluated
by the time integral of $\rho(t)$ for $\lambda\ge 15$ cm$^{-1}$.
The resulting trapping time  and population fluxes correspond to full quantum dynamics.

\subsection{Trapping Time}
\label{sec05a}

Here we  study the quantum-classical comparison with the change of $\lambda$ by fixing
$T=300$ K and $1/D=50$ fs. The dependence on other parameters
can be  explored similarly. Figure~\ref{fig03} presents
the results of the trapping time computed using both quantum dynamics ($\langle t\rangle_Q$) and
hopping kinetics ($\langle t\rangle_C$) with two distinct initial conditions in the seven-site FMO model.
We observe that $\langle t\rangle_Q$ and $\langle t\rangle_C$ are still close,
0.1\% and 11\% under the physiological reorganization energy $\lambda=35$ cm$^{-1}$, for the initial conditions I and II,
respectively.
Full quantum dynamics does not necessarily lead to a faster energy transfer process.
With initial condition I, $\langle t\rangle_Q$ is smaller than $\langle t\rangle_C$
for $\lambda>34$ cm$^{-1}$; with initial condition II,
$\langle t\rangle_Q$ is larger than $\langle t\rangle_C$ in the complete range
of $\lambda$.
The latter behavior is consistent with a recent calculation of energy transfer rate in the two-site system~\cite{Ishizaki2009:JCP1}.

To understand our quantum-classical comparison, we need to clarify higher-order
corrections to the leading-order hopping kinetics in a quantum network
at finite temperatures. (1) As shown in the HSR model,
the first effect arises from multi-site quantum coherence, which facilitates the barrier-crossing energy transfer.
This effect is beyond the Fermi golden rule rate (or the second-order truncation in a general manner) and arises from direct interconversion of various
off-diagonal elements of the reduced density matrix.
The multi-site quantum coherence, including tunneling, interference, and delocalization, has been 
discussed before~\cite{Cao2009:JPCA,JLWu2010:NJP,Ishizaki2009:PNAS,Hoyer2011:ArXiv} and is now identified
in the quantum-classical comparison. In a sense, our systematic expansion and flux network analysis
provides quantitative measures to describe these effects. To distinguish from the two-site quantum coherence
inherent in the Fermi's golden rule rate, we will use the concept of the multi-site quantum coherence to define nontrivial
quantum effects throughout this paper.
(2) The second effect comes from the non-Markovian bath. Without instantaneous bath relaxation,
the system is retained in its previous state, slowing down energy transfer.
In the second-order truncation method, a non-Markovian memory kernel can be still extracted, but the
underlying Born approximation cannot capture the full contribution of bath relaxation. Following the Laplace transformation,
we can prove that the results of $\tau$ and $\langle t\rangle$ are the same,
calculated by Fermi's golden rule rate or the second-order non-Markovian master equation in the local basis.
Hence, our comparison distinguishes higher-order effects of bath relaxation excluding the Born approximation.
A systematic approach of including bath relaxation in electron transfer has been shown in
Ref.~\cite{Cao2000:JCP2}.
 (3) Other than the above two effects, a more subtle quantum-classical difference
results from the steady-state population distribution. Without trapping, hopping kinetics
always imposes $P_n(t\rightarrow \infty)\propto \exp(-\bt \vare_n)$. Instead,
the quantum steady-state distribution is evaluated by $\rho(t\rightarrow\infty) \propto \Tr_B\{\exp(-\bt H_\mathrm{tot})\}$,
which arises from the rigorous quantum Boltzmann distribution with the consideration of both system and bath~\cite{Moix2012:}.
The steady-state population of the lowest energy trap site can be decreased, implying
a lower probability of energy being trapped (energy transfer efficiency).
The interplay of these three effects suggests that
quantum energy transfer can be much more complicated, compared to
the leading-order hopping kinetics. The identification of various higher-order effects
for the trapping time and transfer efficiency will be
studied in our forthcoming papers.

\subsection{Flux Network}
\label{sec05b}

For the physiological condition of $\lambda = 35$ cm$^{-1}$~\cite{Ishizaki2009:PNAS}, we compare the quantum and classical
flux networks, as shown in Fig.~\ref{fig04}. Similar to the study of the HSR model,
the flux network analysis clarifies the contribution of higher-order corrections, especially multi-site quantum coherence.
In the seven-site FMO model, with the barrier-crossing pathway under initial condition I,
the quantum and classical trapping times are nearly the same but the detailed flux networks
can be quite different: $6\rightarrow (5, 7)\rightarrow 4 \rightarrow 3$ is the major path with a ratio
of $F^{C}_{34} = 75\%$ in `classical' hopping kinetics, whereas
$3\leftarrow 1\rightarrow 2 \rightarrow 3$ dominates in quantum dynamics
with $F^{Q}_{31}+F^{Q}_{32} = 71\%$. The branching probability from BChl 1 to BChl 3 differs
by three times, $F^{Q}_{31}/F^{C}_{31}= 3.06$, and
the overall quantum-classical flux difference is $\chi_F = 32\%$.
The switch of the major energy transfer path results from
the quantum tunneling effect of multi-site quantum coherence. With the downhill pathway
under initial condition II, the flux network structure is the same for both hopping kinetics and quantum dynamics,
with a smaller difference of $\chi_F = 7\%$.

\subsection{Robustness Against the Removal of BChl 4}
\label{sec05c}

To complete the quantum-classical comparison for our quantum dynamic network,
we study the stability of FMO after the removal of BChl 4. As shown in Fig.~\ref{fig03}, the change of
the trapping time $\langle t\rangle$ is small enough to sustain a highly efficient energy transfer.
Consistent with the HSR model, quantum dynamics always leads to a smaller
trapping time than hopping kinetics. This behavior can be interpreted by replacing the removal
of BChl 4 with an  infinite energy barrier, $\vare(\mathrm{BChl~ 4})=\infty$, so that
the quantum tunneling effect is very important. As shown by the flux network analysis,
BChl 4 is no longer the bottleneck site under initial condition I, and the trapping time
$\langle t\rangle_Q$ is unaffected by the removal of BChl 4 over a broad range of
reorganization energy. The stability analysis thus reflects
the multi-site quantum coherence feature of the original energy transfer network.

\section{Conclusions and Discussion}
\label{sec06}

In this paper, we continue our investigation of efficient energy
transfer in light-harvesting systems and compare the prediction of
the trapping time and the population flux calculated by full quantum
dynamics and by `classical' hopping kinetics. The classical white
noise (the HSR model) and the quantum Debye noise are used to model
the protein environment. The quantum dynamics under the quantum
Debye noise is solved by the hierarchy equation, which compares well
to the GBR equation used in our first paper~\cite{JLWu2010:NJP}.
Relative to the rigorous results of quantum dynamics, hopping
kinetics is consistently calculated by Fermi's golden rule rate. In
principle, full quantum dynamics can be mapped to an equivalent
kinetic network of population transfer by a systematic expansion. We
have extended the systematic expansion from the HSR
model~\cite{Cao2009:JPCA} to a general quantum dynamic network,
which will be shown in a forthcoming paper~\cite{JLWu2011:NIBA}. In
this mapping, the leading-order hopping rate is equivalent to
Fermi's golden rule (i.e., the Forster rate for the dipole-dipole
interaction) and is taken as the `classical' hopping limit.
Therefore, our quantum-classical comparison is capable of
systematically illustrating nontrivial quantum effects using
higher-order corrections beyond the second-order truncation with the
Born approximation. Our result is different from a previous approach
based on factorizing the Liouville operator of various dynamic
processes~\cite{Rebentrost2009:JPCB}. In the HSR model, higher-order
effects originate purely from multi-site quantum coherence (direct
interconversion of off-diagonal density matrix
elements)~\cite{Cao2009:JPCA}. For a quantum bath model such as the
Debye spectral density, there exist additional contributions from
bath relaxation (non-Markovianity excluding the Born
approximation)~\cite{Cao2000:JCP2} and the finite temperature
effect~\cite{JLWu2011:NIBA}.

Our investigation of the average trapping time demonstrates that
hopping kinetics compares well with full quantum dynamics, and that
the Forster rate can reliably predict optimal energy transfer in
the seven-site FMO model. Two initial conditions, BChl 1 and BChl 6, are used in our
study, while the initial condition is shifted to
BChl 8 in the eight-site model~\cite{Moix2011}.
For both the HSR model and the Debye spectral density, the quantum and
classical trapping times are close, differing a few percentages over
a broad range of parameters around the optimal and physiological
conditions. 
In the local site basis, the energy difference ($\Delta$) between two neighboring
levels is several times larger than their electronic coupling ($J$).
The study in our first paper suggests that an intermediate
dissipation strength ($\Ga$), is necessary for the optimal energy
transfer. A crude estimation on the kinetic expansion parameter,
$J^2/(\Delta^2+\Ga^2)<1$, indicates that quantum-classical
difference can be treated as a small correction to the leading-order
`classical' kinetics in the overall dynamic behavior, i.e., the
trapping time and the transfer efficiency.

However, nontrivial quantum coherent effects can be fundamentally
important in the detailed behavior of the energy transfer process. A
better measurement of nontrivial quantum effects involves
off-diagonal coherence ($\rho_{mn(\neq m)}$) of the reduced density
matrix. In this paper, we propose a new measure, the integrated
population flux (or equivalently the branching probability), which
is defined using the decay time of the off-diagonal density matrix
elements $\rho_{mn}$ in full quantum dynamics. The flux thus defined
obeys the conservation law and is a unique measure of
non-equilibrium energy flow in quantum networks. Through the flux
network analysis, we are able to extract the two major energy
transfer pathways, $1\rightarrow 2 \rightarrow 3$ (path A) and
$6\rightarrow (5, 7)\rightarrow 4 \rightarrow 3$ (path B), in the seven-site FMO model.
Here the energy transfer through path A crosses a barrier at BChl 2; path
B is a downhill structure, thus becoming the dominant pathway in
hopping kinetics. With the initial population at BChl 1 and the
physiological condition of the Debye spectral density for the bath,
the quantum tunneling effect switches the dominant pathway from path
B to path A. The trapping time and flux network with the initial
population at BChl 6 are much less affected due to its downhill
structure. The quantum-classical comparison of the flux network thus
characterizes multi-site quantum coherence for various network
structures, and this coherence becomes more pronounced with the
decrease of temperature.
As discussed in Appendix.~\ref{app3},
the two-pathway energy transfer structure can be found
in the eight-site FMO model but the weight of the two pathways
is changed due to the change of the Hamiltonian and the presence of the eighth BChl.

Using the leading-order `classical' kinetics, we present a simple
estimation on the stability of energy transfer against the change of
internal and external parameters. The time scale separation of
energy trap and decay processes, $k_d\langle t\rangle\sim 0.01$, is
a key factor for FMO. Based on the estimation of hopping rate, a
noticeable change in the transfer efficiency requires a dramatic
change in the trapping time, which in turn requires one or two
orders of magnitude change in various parameters. To induce a
permanent damage in FMO, we remove the bottleneck site, BChl 4, and
explore the modified trapping time in both quantum dynamics and
`classical' kinetics. We observe that the multiple pathways help FMO
sustain a less dramatic change in the trapping time, thus ensuring
the robustness of quantum energy transfer.

Our analysis is based on a physically-motivated kinetic mapping of
quantum dynamics. The quantum coherence has been discussed in the
framework of the long-lived quantum beat and
entanglement~\cite{Sarovar2010:NaturePhys}. Complementary to these
studies, we provide a quantitative measurement rather than a
qualitative description for nontrivial higher-order quantum effects.
Using the two-pathway FMO as our model system, we reveal the
contribution of multi-site quantum coherence and its dependence on
the pathway structure. Our approach can be easily applied to other
light-harvesting systems and artificial devices. Specifically,
multi-site quantum coherence can lead to various phenomena, e.g.,
quantum interference between various energy transfer pathways,
quantum phase modulation of a closed transfer loop, and long-range
energy exchange by quantum tunneling~\cite{Cao2009:JPCA}. To
systematically study these nontrivial quantum behaviors as well as
the bath relaxation effect~\cite{Cao2000:JCP2}, we need to develop a
more detailed partition procedure based on our kinetic mapping
technique~\cite{JLWu2011:NIBA}. Our study is limited in the initial
condition with pure populations, and a more general case with
initial coherence will be extended in the future. For the FMO
system, different Hamiltonian models have been applied in
theoretical and experimental
studies~\cite{Adolphs2006:BPJ,SchmidtamBusch2011:JPCL}. The
variation of the Hamiltonian will lead to different results, e.g.,
the  site energy of the additional eighth BChl can be optimized
close to the experimental value~\cite{Moix2011}, but our
quantum-classical comparison strategy is applicable in general to
these new models. As a demonstration,
the calculation of the eight-site model is summarized in Appendix~\ref{app3},
where the generality of our methodology should be not confused
with model-dependent results.

\acknowledgments This work was supported by grants from the National
Science Foundation (Grant CHE-1112825), DARPA (Grant
N66001-10-1-4063). Jianshu Cao and late Bob Silbey were partly
supported by the Center for Excitonics, an Energy Frontier Research
Center funded by the US Department of Energy, Office of Basic Energy
Sciences (Grant DE-SC0001088). JW acknowledges partial support from
the Fundamental Research Funds for the Central Universities in China
(Grant 2011QNA3005) and the National Science Foundation of China
(Grant 21173185).

\appendix

\section{Quantum Effective Kinetic Rate Matrix}
\label{app1}
In this appendix, we will provide a formal approach of kinetic mapping,
and prove that the leading-order hopping rate of the HSR model
can be recovered from Fermi's golden rule rate with a classical white noise.

For a kinetic network satisfying the master equation in Eq.~(\ref{eq11new}),
 the integrated residence time is calculated by the Laplace transform
\be
\hat{P}(z=0) = \left[K+K_t\right]^{-1} P(t=0),
\label{app_eq001}
\ee
where $\hat{P}(z) = \int_0^\infty dt e^{-zt} P(t)$ and $\tau_n=[\hat{P}(z=0)]_n$.
For a quantum dynamic network, we will generate its kinetic mapping
by the constraint of the same integrated residence time $\hat{P}(z=0)$.
The Liouville equation, $\dot{\rho}(t) = - \mathcalL \rho(t)$, is rewritten as
\be
\dot{\rho}_P(t) =&& - \mathcalL_{\sys; PC} \rho_C(t) -\mathcalL_{\trap; P} \rho_P(t) \no \\
\dot{\rho}_C(t) =&& - \mathcalL_{\sys; CP} \rho_P(t) -\left[\mathcalL_{\sys; C}+\mathcalL_{\dissp; C}+\mathcalL_{\trap; C}\right] \rho_C(t)
\label{app_eq002}
\ee
where the indices $P$ and $C$ represent diagonal population elements and off-diagonal coherence elements of the density matrix
in the local site basis.
The reduced density matrix $\rho$ is separated into two block elements, $\rho_P$ and $\rho_C$.
Each Liouville superoperator ($\mathcalL_{\sys}$, $\mathcalL_{\trap}$, and $\mathcalL_{\dissp}$),
is also separated into the block-matrix form.
In the HSR model, the dissipation Liouville superoperator
is described the pure dephasing constant $\Ga^\ast$, i.e., $\mathcalL_{\dissp; C}=\Ga^\ast$.
In general, the influence of $\mathcalL_{\dissp; C}$ can be described by a time-convolution form,
i.e., $\mathcalL_{\dissp; C}\rho_C(t) = \int_0^t M(t-\tau)\rho_C(\tau)d\tau$.
The time-nonlocal dissipation kernel $M(t)$ can be formally expressed in terms of projection operators.
Applying the Laplace transform, we obtain a closed form for the population vector,
\be
\hat{\rho}_P(z) &=& \left\{z-\mathcalL_{\sys; PC}\left[z+\mathcalL_{\sys; C}+\hat{M}(z=0)+\mathcalL_{\trap;C}\right]^{-1}
\mathcalL_{\sys;CP}+\mathcalL_{\trap; P} \right\}^{-1}  \no \\
&&\times \rho_P(t=0),
\ee
where $\hat{\rho}_P(z)$ 
and $\hat{M}(z)$ are the population vector and the disspation Liouville superoperator in the Laplace $z$-domain,
respectively.
To derive the above equation, we presume zero initial quantum coherence, $\rho_C(t=0)=0$.
With two identities, $P=\rho_P$ and $K_t=\mathcalL_{\trap; t}$,
the integrated residence time vector from the full quantum dynamics is given by
\be
\hat{P}(z=0) = \left\{-\mathcalL_{\sys; PC}\left[\mathcalL_{\sys; C}+\hat{M}(z=0)+\mathcalL_{\trap;C}\right]^{-1}
\mathcalL_{\sys;CP}+K_t \right\}^{-1}P(t=0).
\label{app_eq004}
\ee
Comparing Eq.~(\ref{app_eq001}) and Eq.~(\ref{app_eq004}), we obtain the quantum kinetic rate matrix as
\be
K^Q = -\mathcalL_{\sys; PC}\left[\mathcalL_{\sys; C}+\hat{M}(z=0)+\mathcalL_{\trap;C}\right]^{-1}
\mathcalL_{\sys;CP},
\label{app_eq005}
\ee
which includes the leading-order 'classical' hopping rates and higher-order corrections from multi-site quantum coherence.
The effective quantum rate is given by $k^Q_{mn(\neq m)} = -[K^Q]_{mn}$.

In the leading order, we ignore the off-diagonal elements of $\mathcalL_{\sys; C}$, i.e.,
$\mathcalL_{\sys; C}\approx \mathcalL^{(0)}_{\sys; C}\rightarrow i\Delta_{mn}$.
With the explicit form of $\mathcalL_{\sys}$ in Eq.~(\ref{eq01a}), this simplification allows us to obtain
the `classical' hopping rate $k^C_{mn}~(\neq k^Q_{mn})$ in Eq.~(\ref{eq10}) for the HSR model.
In a forthcoming paper~\cite{JLWu2011:NIBA}, we will demonstrate the mapping procedure for the general quantum dynamic network.
The leading-order hopping rate is given by Fermi's golden rule rate in Eq.~(\ref{eq13a}).
For a classical white noise described by  $J(\omega) = (\bt\Ga^\ast/2\pi)\omega$, we ignore the constant imaginary part of $g(t)$
and arrive at
\be
g(t) \approx 2 \int_0^\infty d\omega \frac{(\bt\Ga^\ast/2\pi)\omega}{\omega^2} \frac{2}{\bt\omega}(1-\cos\omega t)=\Ga^\ast t
\ee
and
\be
k^C_{m\neq n} =&& 2 |J_{mn}|^2 \Real \int_0^\infty d\tau e^{-(\Ga_{mn}+i \Delta_{mn}) \tau}=2 |J_{mn}|^2\frac{\Ga_{mn}}{\Ga^2_{mn}+\Delta^2_{mn}},
\ee
which recovers the result in the HSR model.

\section {Derivation of Quantum Integrated Population Flux}
\label{app2}

The integrated population flux $F^C_{mn}$  in a classical kinetic network is
defined in Eq.~(\ref{apeq01}). The corresponding quantity, $F^Q_{mn}$, in a quantum kinetic network is similarly defined using
$k^Q$ and $\tau^Q$. Since the quantum effective rate $k^Q$ is difficult to be exactly determined,
we will rewrite $F^Q_{mn}$ using the time integration of coherence.
Applying the Laplace transform to the second equation of Eqs.~(\ref{app_eq002}) and setting the Laplace variable to be zero
($z=0$), we obtain
\be
-\mathcalL_{\sys; CP}\hat{\rho}_P(0) = \left[\mathcalL_{\sys; C}+\hat{M}(z=0)+\mathcalL_{\trap;C}\right]\hat{\rho}_C(0),
\ee
where the condition of zero initial quantum coherence ($\rho_C(t=0)=0$) is used. The above equation be further
rearranged, giving
\be
K^Q \hat{P}(0) = \mathcalL_{\sys; PC}\hat{\rho}_C(0)
\ee
with the help of $K^Q$ defined in Eq.~(\ref{app_eq005}). The $n$-th vector element on the both sides of this equation is
written as
\be
\sum_{m(\neq n)} \left[k^Q_{mn}\tau^Q_n-k^Q_{nm}\tau^Q_m\right] = i \sum_{m(\neq n)} \left(J_{nm}\tau^Q_{mn}-J_{mn}\tau^Q_{nm}\right),
\ee
where the coherence decay time $\tau^Q_{mn}=\int_0^\infty dt \rho_{mn}(t)$ is introduced.
Since the indices, $m$ and $n$, are arbitrary in the above summation, we obtain the quantum population flux as
\be
F^Q_{mn} = k^Q_{mn}\tau^Q_n-k^Q_{nm}\tau^Q_m = 2~ \Imag\left[ J_{mn}\tau^Q_{nm}\right],
\label{app_eq020n}
\ee
with two identities, $J_{nm} = J^\ast_{mn}$ and $\tau^Q_{nm} = [\tau^Q_{mn}]^\ast$.

\section{Summary of Quantum-Classical Comparison and Flux Network Analysis for the Eight-Site FMO Model}
\label{app3}

Since the last year, the crystal structure of FMO has been revisited and a new eight-site model has been
proposed~\cite{SchmidtamBusch2011:JPCL}. The additional eighth BChl is considered as the initial site 
for energy transfer in FMO. Quantum dynamics in the original seven-site model is modified, leading
to different observations. In the third paper of this series~\cite{Moix2011}, we explained the suppressed oscillation in
the eight-site FMO model and proposed an optimal equally-spaced ladder structure for the $8\rightarrow(1,2)\rightarrow 3$ pathway.
Our methodology of quantum-classical comparison and flux network analysis
is however general and model independent. Here we provide a short summary of our calculation for the eight-site model for the
completeness.

Applying the Hamiltonian of the eight-site model from Ref.~\cite{SchmidtamBusch2011:JPCL}, which can be
found in the third paper of this series~\cite{Moix2011}, we perform the quantum (hierarchic equation)
and classical (Fermi's golden rule rate) calculations described in Sec.~\ref{sec05}.
The bath is modeled by the Debye spectral density with the same set of parameters ($\lambda = 35$ cm$^{-1}$, $1/D = 50$ fs and $T=300$ K).
The high-temperature approximation is used to reduce the numerical cost.
The resulting two trapping times are $\langle t\rangle_C = 4.05$ ps and $\langle t\rangle_Q= 4.69$ ps, with the relative
difference of $\sim 14\%$ between quantum dynamics and hopping kinetics. 
Compared to the result of the initial condition at BChl 1 in the seven-site FMO model, the difference
of $\langle t\rangle_Q$ is less than 5\%, indicating the stability of energy transfer efficiency
against the change of the Hamiltonian. 
Next we construct the flux network using both quantum and classical approaches. As shown in Fig.~\ref{fig05},
the two major energy transfer pathways, $8\rightarrow 1\rightarrow 2 \rightarrow 3$ and $6\rightarrow (5, 7)\rightarrow 4 \rightarrow 3$,
are determined by relevant net population flows ($>0.1$), and the overall structure of the flux network is kept the same
under quantum and classical descriptions. The average difference between the two flux networks is $\chi_F\approx 20\%$,
slightly larger than the difference of the trapping time. Compared to the case of the initial condition at BChl 1 in
the seven-site model, the energy transfer in the eight-site model is more dispersed in the network,
and the probabilities of the two pathways are closer ($F_{32}\approx F_{34}$).
A small but unidirectional inter-path energy flow can be observed
from the first pathway,  $8\rightarrow (1, 2) \rightarrow 3$, to the second pathway, $6\rightarrow (5, 7)\rightarrow 4 \rightarrow 3$.
These two phenomena are related to the Hamiltonian used in the eight-site model: 1) the average energy level of the first pathway
is higher than that of the second pathway; 2) the coupling between the trap site, $J_{34}\approx 2 J_{32}$, prefers the second pathway, especially
in the classical description;
3) the additional eighth site allows more chances of inter-path energy flows and further increases the dispersibility. One the other hand,
quantum-classical comparison also reveals a noticeable change in the weights of the two pathways ($\chi_F\approx 20\%$).
Quantum mechanically, the multi-site quantum coherence
allows a long-range direct energy transfer from BChl 8 to BChl 2.
The increase of quantum branch probability from BChl 2 to BChl 3 compared to the classical value results
from the interplay of multi-site quantum coherence and solvent relaxation effect.

In a summary, we apply the quantum-classical comparison and the flux network analysis to a new eight-site FMO model.
The basic two-pathway energy transfer structure can be still observed in the eight-site FMO model. Because of the change in
the Hamiltonian, the detailed results of quantum-classical difference and non-trivial quantum effects are modified accordingly.
Interestingly, energy transfer becomes more dispersed in the eight-site FMO model, which may help the system resist damages.

\newpage
\noindent {\bf Figure Captions}

\begin{figure}[htp]
\includegraphics[width=0.90\columnwidth]{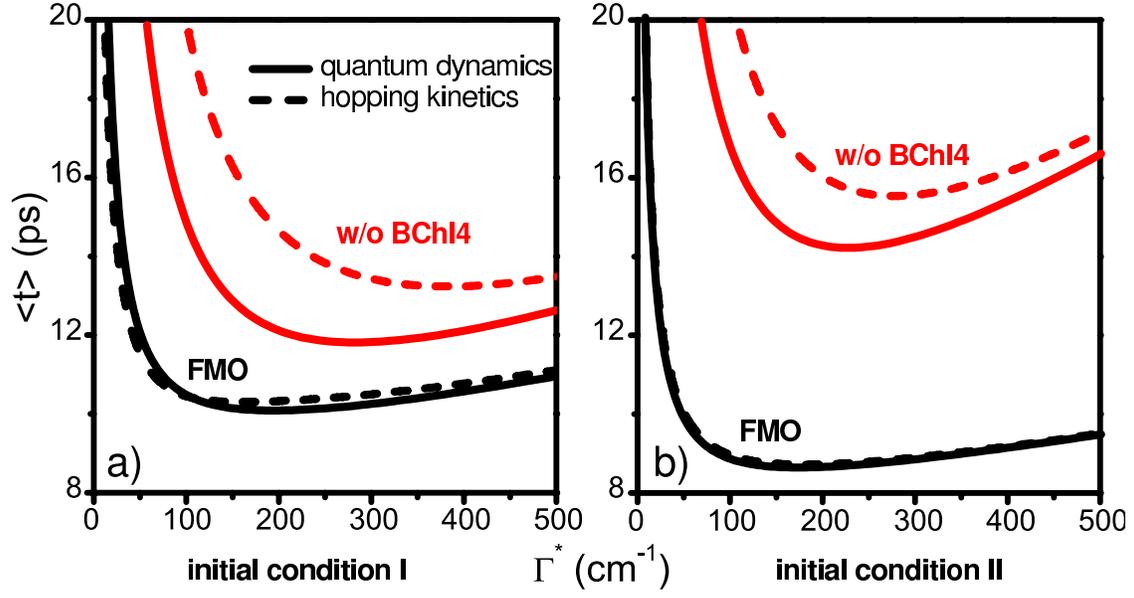}
  \caption{The average trapping time $\langle t\rangle$ vs. the pure dephasing rate $\Ga^\ast$ in the HSR model
of the  seven-site FMO for
  a) the initial population at BChl 1, and b) the initial population at BChl 6. The solid curves are calculated
  from full quantum dynamics, whereas the dashed curves are calculated  from the leading-order `classical' kinetics.
  In each figure, the lower pair of curves correspond to the seven-site FMO model, while the upper pair corresponds
  to the six-site FMO model after the removal of BChl 4.}
  \label{fig01}
\end{figure}

\noindent Fig.~\ref{fig01}: The average trapping time $\langle t\rangle$ vs. the pure dephasing rate $\Ga^\ast$ in the HSR model
of the  seven-site FMO for
  a) the initial population at BChl 1, and b) the initial population at BChl 6. The solid curves are calculated
  from full quantum dynamics, whereas the dashed curves are calculated  from the leading-order `classical' kinetics.
  In each figure, the lower pair of curves correspond to the seven-site FMO model, while the upper pair corresponds
  to the six-site FMO model after the removal of BChl 4.

~

\begin{figure}[htp]
\includegraphics[width=1\columnwidth]{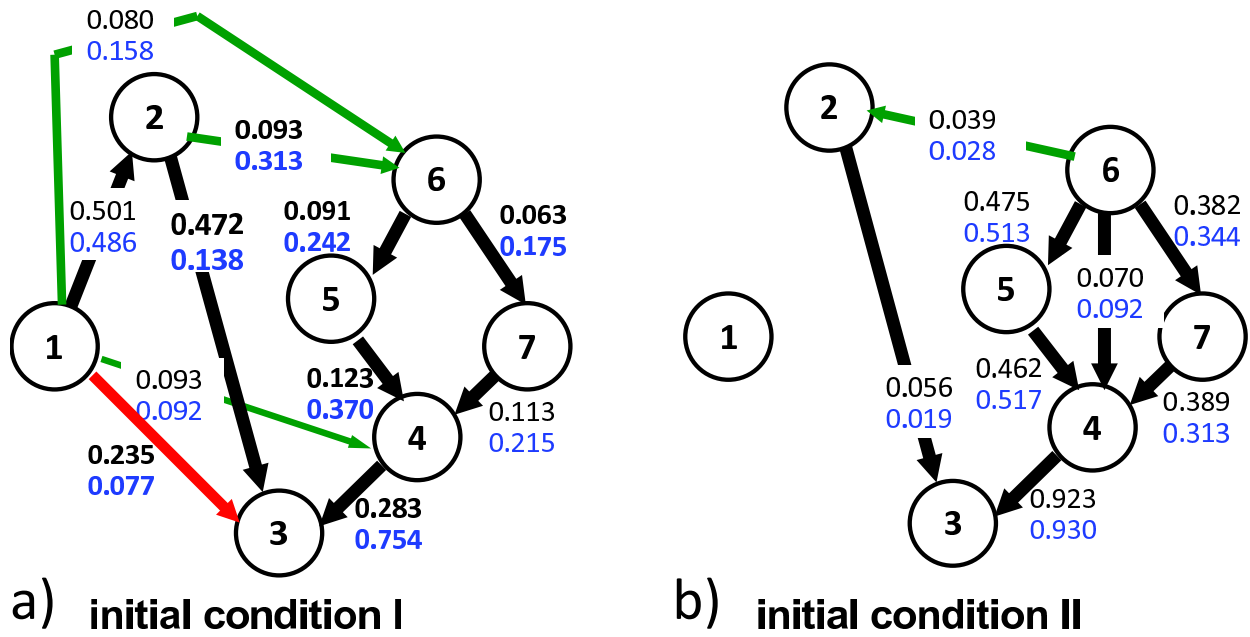}
  \caption{The flux networks of FMO under the optimal pure dephasing rate
  in the HSR model of the  seven-site FMO
 for a) initial population at BChl 1 ($\Ga^\ast_{\opt}=175$ cm$^{-1}$),
  and b) initial population at BChl 6 ($\Ga^\ast_{\opt}=195$ cm$^{-1}$). For each population flux,
  the upper number is obtained from full quantum dynamics, whereas the lower number is obtained from
  the leading-order hopping kinetics.}
  \label{fig02}
\end{figure}

\noindent Fig.~\ref{fig02}: The flux networks of FMO under the optimal pure dephasing rate
  in the HSR model of the  seven-site FMO
 for a) initial population at BChl 1 ($\Ga^\ast_{\opt}=175$ cm$^{-1}$),
  and b) initial population at BChl 6 ($\Ga^\ast_{\opt}=195$ cm$^{-1}$). For each population flux,
  the upper number is obtained from full quantum dynamics, whereas the lower number is obtained from
  the leading-order hopping kinetics.

~

\begin{figure}[htp]
\includegraphics[width=0.90\columnwidth]{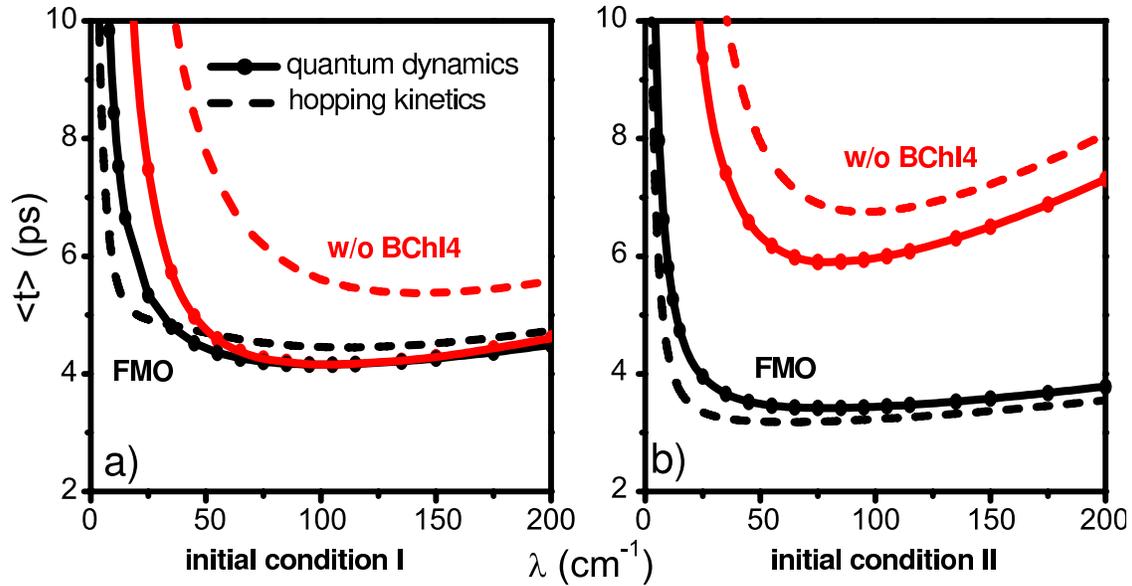}
  \caption{The average trapping time $\langle t\rangle$ of the  seven-site FMO
 vs. the reorganization energy $\lambda$ of the Debye spectral density for
  a) the initial population at BChl 1, and b) the initial population at BChl 6. The other bath parameters are given
  in text.   The solid curves are results
  of full quantum dynamics, whereas the dashed curves results of the leading-order hopping kinetics.
  In each figure, the lower pair of curves (black) correspond to the seven-site FMO model,
  while the upper pair (red) corresponds to the six-site FMO model with the removal of BChl 4. (colored online)}
  \label{fig03}
\end{figure}

\noindent Fig.~\ref{fig03}: The average trapping time $\langle t\rangle$ of the  seven-site FMO
 vs. the reorganization energy $\lambda$ of the Debye spectral density for
  a) the initial population at BChl 1, and b) the initial population at BChl 6. The other bath parameters are given
  in text.   The solid curves are results
  of full quantum dynamics, whereas the dashed curves results of the leading-order hopping kinetics.
  In each figure, the lower pair of curves (black) correspond to the seven-site FMO model,
  while the upper pair (red) corresponds to the six-site FMO model with the removal of BChl 4. (colored online)

~

\begin{figure}[htp]
\includegraphics[width=1\columnwidth]{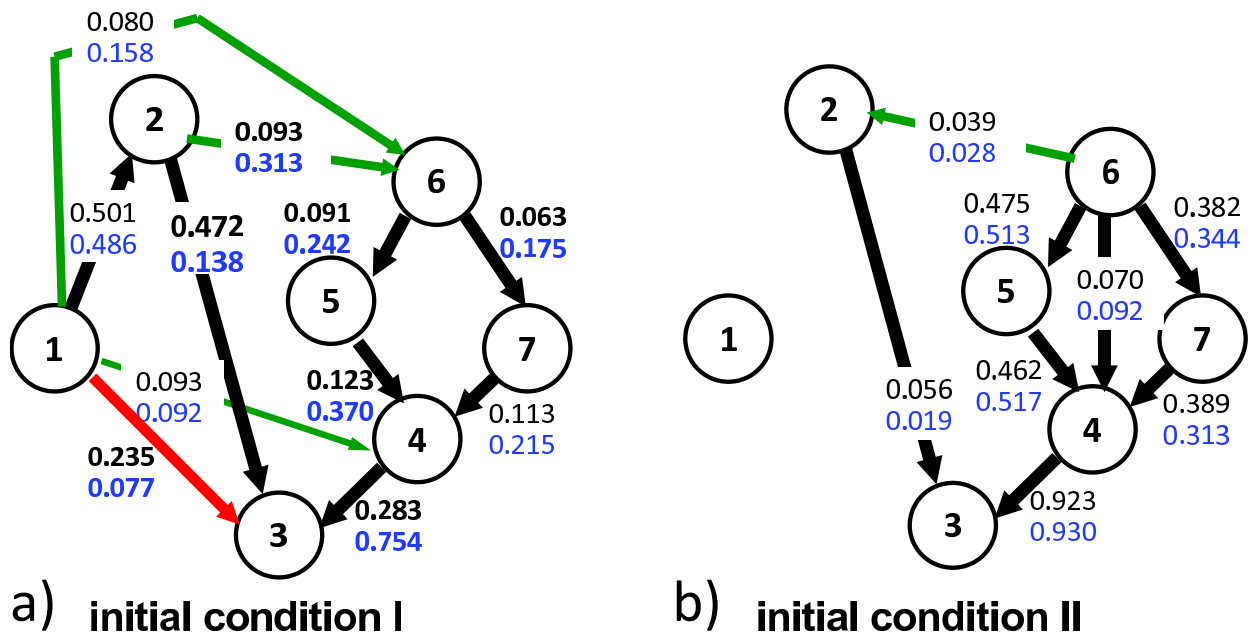}
  \caption{The flux networks of the seven-site FMO under the physiological condition ($\lambda = 35$ cm$^{-1}$, $1/D = 50$ fs, and $T = 300$ K)
  of the Debye spectral density for a) initial population at BChl 1,
  and b) initial population at BChl 6. For each population flux,
  the upper number is obtained using full quantum dynamics whereas the lower number is obtained using
  the leading-order `classical' hopping kinetics.}
  \label{fig04}
\end{figure}

\begin{figure}[htp]
\includegraphics[width=1\columnwidth]{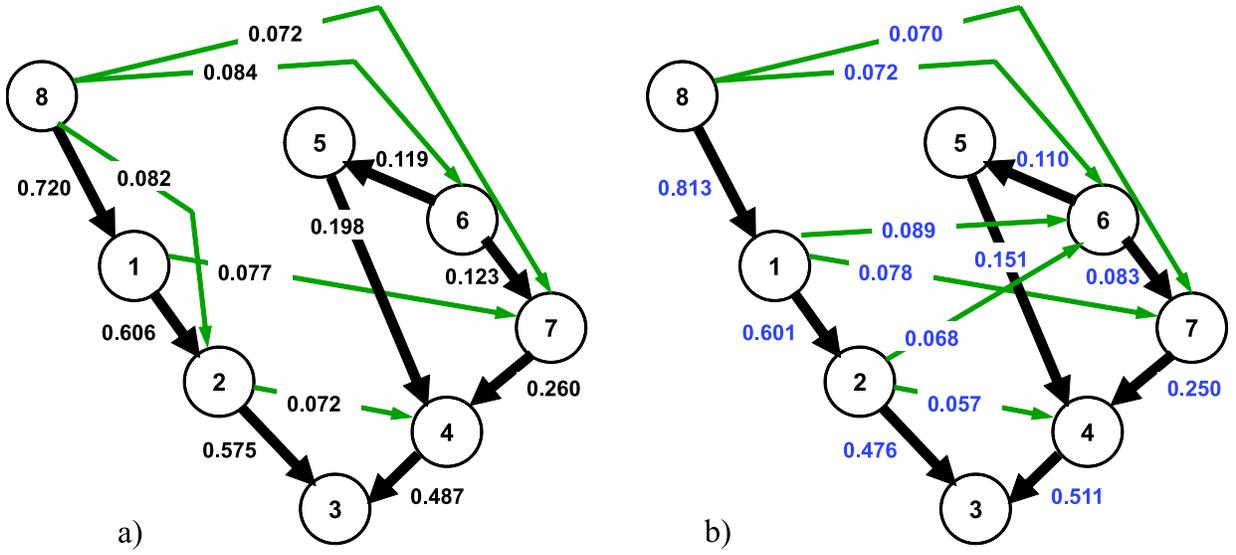}
  \caption{
The flux networks of the eight-site FMO under the physiological condition ($\lambda = 35$ cm$^{-1}$, $1/D = 50$ fs, and $T = 300$ K)
  of the Debye spectral density using a) full quantum dynamics and b) the leading-order classical kinetics.}
  \label{fig05}
  \end{figure}

\noindent Fig.~\ref{fig04}: The flux networks of the seven-site FMO under the physiological condition ($\lambda = 35$ cm$^{-1}$, $1/D = 50$ fs, and $T = 300$ K)
  of the Debye spectral density for a) initial population at BChl 1,
  and b) initial population at BChl 6. For each population flux,
  the upper number is obtained using full quantum dynamics whereas the lower number is obtained using
  the leading-order `classical' hopping kinetics.

~

\noindent Fig.~\ref{fig05}: The flux networks of the eight-site FMO under the physiological condition ($\lambda = 35$ cm$^{-1}$, $1/D = 50$ fs, and $T = 300$ K)
  of the Debye spectral density using a) full quantum dynamics and b) the leading-order classical kinetics.

\newpage

\newpage

\newpage

\newpage

\newpage

\end{document}